\documentclass[conference, final]{IEEEtran}
\usepackage{amsmath,amssymb,amsfonts}
\usepackage{algorithmic}
\usepackage{graphicx}
\usepackage{textcomp}
\usepackage{xcolor}
\usepackage{physics}
\usepackage{hyperref}
\usepackage{booktabs}
\usepackage{braket}
\usepackage{microtype}

\graphicspath{ {./figures/} }

\def\BibTeX{{\rm B\kern-.05em{\sc i\kern-.025em b}\kern-.08em
    T\kern-.1667em\lower.7ex\hbox{E}\kern-.125emX}}

\newcommand{\bigO}{\mathcal{O}}

\usepackage{tikz}

\newcommand\copyrighttext{%
  \footnotesize \textcopyright 2023 IEEE. Personal use of this material is permitted.
  Permission from IEEE must be obtained for all other uses, in any current or future
  media, including reprinting/republishing this material for advertising or promotional purposes, creating new collective works, for resale or redistribution to servers or lists, or reuse of any copyrighted component of this work in other works. 
  DOI: \href{https://doi.org/10.1109/QCE57702.2023.00075}{10.1109/QCE57702.2023.00075}
  }
\newcommand\copyrightnotice{%
\begin{tikzpicture}[remember picture,overlay]
\node[anchor=south,yshift=10pt] at (current page.south) {\fbox{\parbox{\dimexpr\textwidth-\fboxsep-\fboxrule\relax}{\copyrighttext}}};
\end{tikzpicture}%
}

\begin{document}

\title{Semisupervised Anomaly Detection using Support Vector Regression with Quantum Kernel}

\author{\IEEEauthorblockN{Kilian Tscharke, Sebastian Issel, Pascal Debus}
\IEEEauthorblockA{\textit{Quantum Security Technologies} \\
\textit{Fraunhofer Institute for Applied and Integrated Security}\\
Garching near Munich, Germany \\
\{kilian.tscharke, sebastian.issel, pascal.debus\}@aisec.fraunhofer.de}
}

\maketitle
\copyrightnotice

\begin{abstract} 
Anomaly detection, an important task in data analysis, involves identifying observations or events that deviate in some way from the rest of the data.
Machine learning techniques have shown success in automating this process by detecting hidden patterns and deviations in large-scale data.
The potential of quantum computing for machine learning has been widely recognized, leading to extensive research efforts to develop suitable quantum machine learning (QML) algorithms.
In particular, the search for QML algorithms for near-term NISQ devices is in full swing.
However, NISQ devices pose additional challenges due to their limited qubit coherence times, low number of qubits, and high error rates.
Kernel methods based on quantum kernel estimation have emerged as a promising approach to QML on NISQ devices, offering theoretical guarantees, versatility, and compatibility with NISQ constraints.
Especially support vector machines (SVM) utilizing quantum kernel estimation have shown success in various supervised learning tasks.
However, in the context of anomaly detection, semisupervised learning is of great relevance, and yet there is limited research published in this area.

This paper introduces an approach to semisupervised anomaly detection based on the reconstruction loss of a support vector regression (SVR) with quantum kernel.
This novel model is an alternative to the variational quantum and quantum kernel one-class classifiers, and is compared to a quantum autoencoder as quantum baseline and a SVR with radial-basis-function (RBF) kernel as well as a classical autoencoder as classical baselines.
The models are benchmarked extensively on 10 real-world anomaly detection data sets and one toy data set and it is shown that our SVR model with quantum kernel performs better than the SVR with RBF kernel as well as all other models, achieving highest mean AUC over all data sets.
In addition, our QSVR outperforms the quantum autoencoder on 9 out of 11 data sets. 
\end{abstract}


\section{Introduction}
In data analysis, anomaly detection refers to the often \textit{unsupervised} task of identifying patterns, observations, or events that exhibit significant deviations from the expected or normal behavior within a dataset \cite{Ruff2020}.
These outliers or novelties may represent rare events, errors, fraud, or other interesting phenomena that differ from the majority of the data.
Anomaly detection has wide-ranging applications in fields such as finance, healthcare, industrial monitoring, as well as in cybersecurity and fraud detection, where the detection of unusual occurrences or behaviors can have significant implications for identifying potential security threats, preventing data breaches, detecting malicious activities, and safeguarding critical systems and assets from unauthorized access or malicious activities \cite{Ruff2020}.

Due to the increasing availability of vast amounts of data, traditional rule-based methods may struggle to effectively handle the scale and complexity of modern datasets.
As a result, machine learning has become a significant approach in anomaly detection, leveraging algorithms and models to automatically identify patterns and anomalies in large-scale data.
These approaches are capable of processing and analyzing vast amounts of information to uncover hidden patterns and detect deviations from expected behavior and have shown success in analyzing and identifying anomalies from massive datasets \cite{Thudumu2020}.
However, even traditional machine learning methods have limitations in terms of computational power and efficiency when dealing with large datasets with high-dimensional feature spaces \cite{Thudumu2020}.

Quantum machine learning may offer a promising approach for overcoming challenges in processing and analyzing such large datasets. The unique properties of quantum systems like superposition and the exponential size of the Hilbert space provide the potential to handle larger amounts of information than what is feasible classically.
Moreover, quantum computing may offer speedup and energy-saving benefits due to the efficient calculations enabled by superposition and entanglement.
Additionally, research suggests that quantum machine learning methods can achieve effective training with fewer samples compared to classical models, and in some cases, even a single sample may be sufficient for predicting quantum data \cite{Kottmann2021}.
Finally, quantum machine learning is believed to be able to work on today's NISQ devices \cite{Havlicek2019, Perdomo-Ortiz_2018}.
These advantages make quantum machine learning a compelling avenue for advancing anomaly detection techniques.

The successful solution of regression problems using SVR can be limited when dealing with large feature spaces and computationally expensive kernel functions \cite{Havlicek2019}.
In such cases, quantum algorithms could potentially offer computational speed-ups by exploiting the exponentially large quantum state space through controllable entanglement and interference.
This motivates the use of Quantum Support Vector Regression (QSVR) as a potential solution to overcome these limitations and leverage the power of quantum computing in regression tasks involving large feature spaces.

The remainder of this paper is structured as follows: The next subsection gives an overview over the related work in the field of quantum support vector machines and quantum autoencoders.
In the final part of the introduction, we will explain the contributions of our paper.
In the following background section, the fundamentals of anomaly detection, quantum machine learning on NISQ devices, (quantum) support vector regression and the (quantum) autoencoder are introduced.
Next, in the methods section the implementation of the models and the datasets is described.
In the results and discussion we show the results of our experiments and analyse them.
Finally, the outlook gives future research directions.

\subsection{Related Work}
Rebentrost, Mohseni and Lloyd \cite{Rebentrost2014} showed in 2014 that a SVM can be implemented on a fault-tolerant quantum computer with complexity logarithmic in the size of the vectors and the number of training samples. 
In 2019, Havlicek et al. \cite{Havlicek2019} proposed and experimentally implemented a SVM with quantum kernel for binary classification on two qubits of a NISQ device.
Since then, QSVMs have been implemented for different applications, e.g. remote sensing image classification \cite{Delilbasic2021}, mental health treatment prediction \cite{Ahmad2021}, breast cancer prediction \cite{Mafu2021} and proton collision at the Large Hadron Collider \cite{LHC_QSVM}. 
In 2022, Kyriienko and Magnusson \cite{quantum_one_class_SVM} performed unsupervised fraud detection with a quantum kernel model based on the one-class SVM.
They performed simulations for systems of up to 20 qubits, and showed that their quantum kernel model achieved better average precision than the classical one-class SVM.

Schuld \cite{Schuld2021_qml_are_kernel} finds that the mathematical structure of supervised quantum machine learning algorithms is closely related to kernel methods.
The algorithms analyse data in high-dimensional Hilbert spaces to which we only have access through inner products revealed by measurements.
Furthermore she states that a lot of near-term and fault-tolerant quantum models can be replaced by a general SVM whose kernel computes distances between data-encoding quantum states, and that kernel-based training is guaranteed to find better or equally good quantum models than variational circuit training.
Finally, she concludes that the data encoding is  key to setting quantum models apart from classical ML models.

In 2017, Romero, Olson and Aspuru-Guzik \cite{Romero2016} introduced a quantum autoencoder (QAE) trained via classical optimization for efficient compression of quantum data.
They applied the model to compress ground states of the Hubbard model and molecular Hamiltonians.
Bravo-Pieto \cite{Bravo-Pieto2021} presented an enhanced feature QAE that is capable of compressing quantum states with higher fidelity than standard QAEs.
Kottmann, Metz, Fraxanet and Baldelli \cite{Kottmann2021} proposed a QAE for detecting anomalies in ground states of quantum many-body Hamiltonians.
Unlike Bravo-Pieto's autoencoder, their model does not need the decoder to calculate the fidelity between input and output states, resulting in a simpler architecture.
Instead, the ability of the encoder to efficiently compress normal samples in latent space is used for the classification.
The model consists of a variational quantum eigensolver that prepares the ground states and a parameterized unitary whose parameters are optimized during training.
Finally, a subset of the qubits, the so called "trash qubits", is measured and the loss is calculated from the result.
The training of the model is done by a classical feedback loop, where the calculation of the loss is the only part performed on a quantum computer.
They demonstrated the feasibility of generating the phase diagram of a quantum system using just a single training sample, without requiring any prior knowledge of the system, such as the order parameter.
In 2022, Ngairangbam, Spannowsky and Takeuchi used a QAE for anomaly detection in high-energy physics at the Large Hadron Collider.
The maximum input dimension of the autoencoders was six, and their simulated QAE performed better than a classical autoencoder.

Recently, Park, Huh and Park \cite{var_quant_one_class_classifier} introduced a variational quantum one-class classifier and compared it to various classical models, among others to an autoencoder and a one-class SVM.
They showed that the performance of their model is comparable to that of a one-class SVM, and even better than the autoencoder in most cases under similar training conditions.
Their model also has the advantage that the parameters grow only logarithmically with the data size.

\subsection{Contributions} 
In this paper, we propose a novel approach for semisupervised anomaly detection using support vector regression with a quantum kernel.
Our model is evaluated extensively using 5 metrics on 10 real-world data sets commonly used for anomaly detection tasks, and a synthetic toy data set.
Our model is compared to a quantum autoencoder based on \cite{Kottmann2021}, as well as to the classical counterparts for both quantum models.
We show that our model can compete with the classical models and even outperforms the quantum autoencoder.
This comparative analysis establishes a benchmark for real-world applications of future quantum anomaly detection (QAD) models.

\section{Background}
\subsection{Anomaly Detection}
Let $\mathcal{X}$ be the data space and $x \in \mathcal{X}$ the feature vector of a single sample.
Let $\mathcal{Y}$ denote the target variable space and $y \in \mathcal{Y}$ the target variable of a single sample \cite{Sperl2021}.
Let further $\Theta$ denote the space of model parameters.
The general task of \textit{supervised} machine learning is to train a parameterized model $f_\theta: \mathcal{X} \times \Theta\to\mathcal{Y}$ such that it approximates a mapping between input $x$ and output $\hat{y}$ based on the learned parameters $\theta$ \cite{Cunningham2008_supervised}, as described in \eqref{eq_output_ad}.
During training, the parameters $\theta$ are optimized such that the loss $\mathcal{L}$ between the output $\hat{y}$ and the target variable $y$ is minimized, as in \eqref{eq_loss_ad}.
\begin{align}
    \hat{y} = f_\theta(x) \label{eq_output_ad} \\
    \min_\theta \mathcal{L}(y, \hat{y}) \label{eq_loss_ad}
\end{align}
Anomaly detection, however, can also be a \textit{semisupervised} or \textit{unsupervised} learning task \cite{Chandula2009}.
In the semisupervised case the training data contains only samples which are assumed to be normal whereas in the unsupervised case, the data has no labels at all.
In addition, in anomaly detection the datasets are usually highly unbalanced, i.e. there are far fewer anomalies than normal instances \cite{Chandula2009}.
These restrictions make semisupervised and unsupervised anomaly detection in general more challenging than supervised learning \cite{Cunningham2008_unsupervised}. 
In semi- or unsupervised anomaly detection, the output $\hat{y}$ of the model often is an anomaly score describing how anomalous the sample is \cite{Chandula2009}.
The anomaly score can for instance be reconstruction-based, i.e. the model is trained to reconstruct the data with the assumption that the model learns the patterns of normal data whereas it is unable to reconstruct anomalies well \cite{huang2022enhancing}.
The decision boundary of whether a given sample is classified as anomalous or not can then be defined by a threshold $\tau$ for the anomaly score:

\begin{align}
    \text{normal if} \: \hat{y} &< \tau  \\
    \text{anomalous if} \: \hat{y} &> \tau \notag
\end{align}

\subsection{QML on NISQ}
\begin{figure}[htbp]
\centerline{\includegraphics[width=0.5\textwidth]{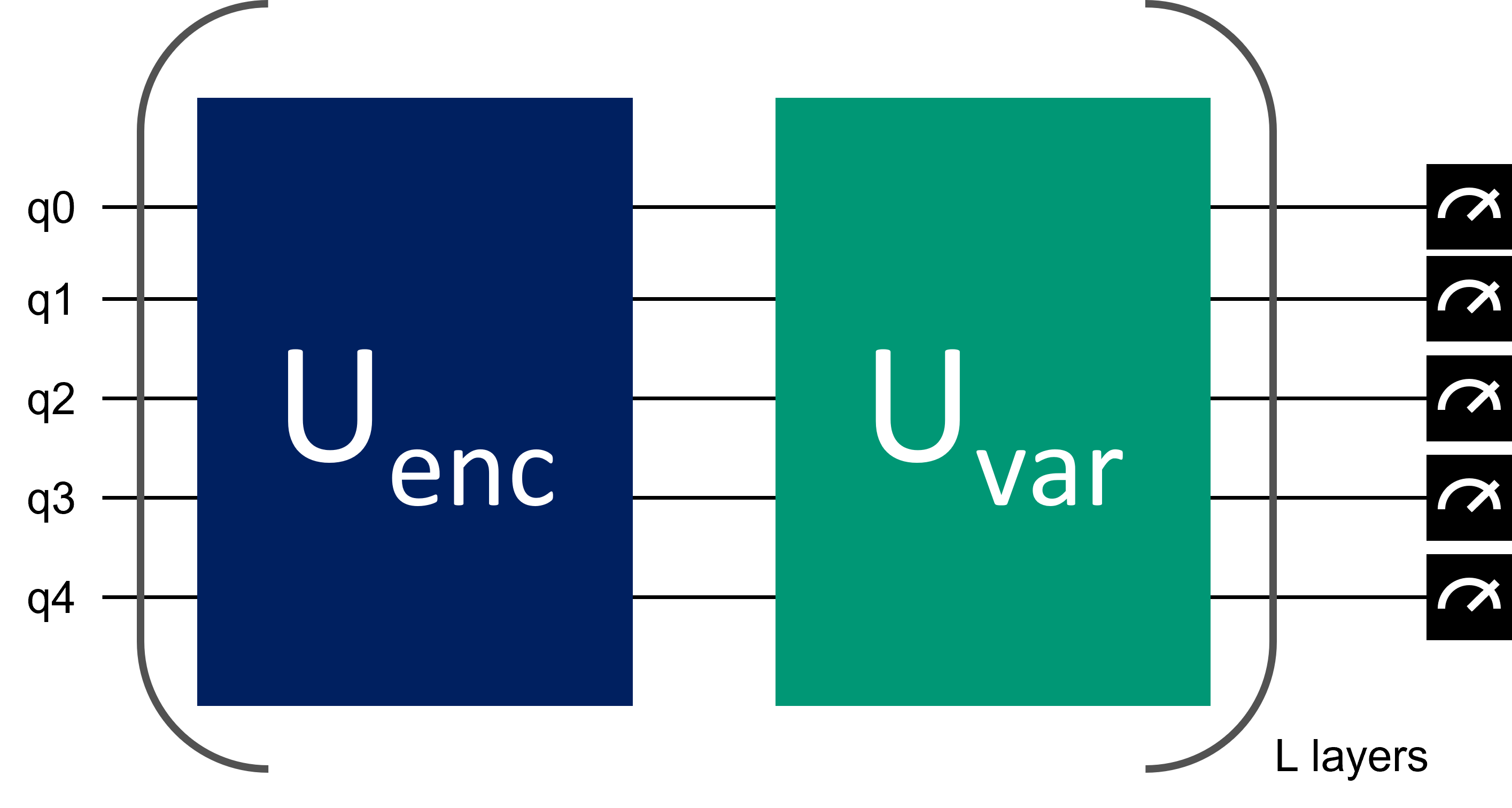}}
\caption{General architecture of a variational model.}
\label{fig_var_circuit}
\end{figure}

\begin{figure}[htbp]
\centerline{\includegraphics[width=.5\textwidth]{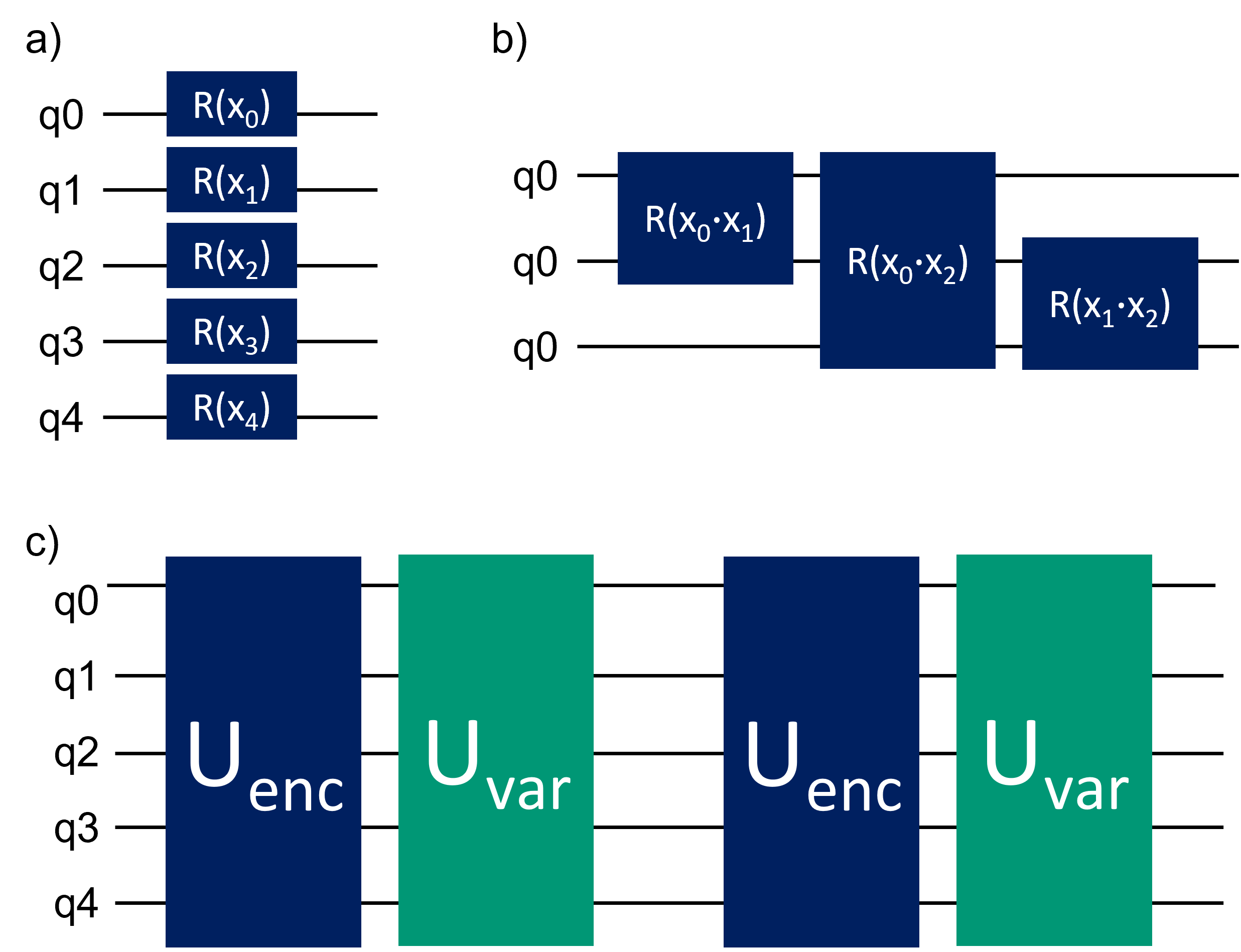}}
\caption{Overview of a) angle encoding, b) IQP-style encoding and c) data re-uploading.}
\label{fig_enc_circuit}
\end{figure}

Quantum machine learning (QML) is an emerging field that harnesses the potential power of quantum computing for enhancing machine learning tasks.
Despite the high error rates, low coherence times and limited number of qubits, today's NISQ devices offer a unique platform for exploring the potential of QML.
Intriguingly, the inherent noise in quantum systems may even be beneficial to QML algorithms, since some classical ML algorithms profit from noise as well \cite{Ciliberto2018}.
Due to its novelty, there are no established algorithms and methods for QML yet, but many authors use variational circuits consisting of a sequence of encoding and variational layers and train the model by minimizing a loss function resulting from the measurement of the circuit \cite{Kottmann2021, Havlicek2019, Romero2016, Guan2021, Schuld2020_effect_of_encoding}.
Since the loss is typically minimized with a classical optimizer like Adam, these models belong to the class of hybrid algorithms.
Fig.~\ref{fig_var_circuit} shows the general architecture of such a variational model. 
In the encoding layer ($U_{enc}$ in Fig.~\ref{fig_var_circuit}) the data is "loaded" into the circuit.
The variational layer ($U_{var}$ in Fig.~\ref{fig_var_circuit}) contains parameterized gates whose parameters are optimized during training of the model.
The sequence of encoding and variational layers is repeated L times.
The encoding can be done e.g. by angle encoding or amplitude encoding.
For \textit{angle encoding}, each (rescaled) feature is the parameter of a rotational gate, as it is shown in Fig.~\ref{fig_enc_circuit} a).
This encoding requires $\bigO(n)$ gates for a feature vector of length $n$ and usually $n$ qubits.
For more advanced versions of angle encoding, a single feature can be encoded into multiple rotation gates and/or IQP-style encoding can be applied, where two-qubit rotational gates are used with the parameters being a product of two individual features, as shown in Fig.~\ref{fig_enc_circuit} b).
In addition, data re-uploading can be used, where a sequence of encoding and variational layers is repeated multiple times, as in Fig.~\ref{fig_enc_circuit} c).

\textit{Amplitude encoding} on the other hand, encodes the feature vector into the amplitudes of the multi qubit state. Here, only $\bigO(log(n))$ qubits are needed, but the number of gates scales potentially exponential with the length of the feature vector \cite{schuld2021machine}.
For example, a feature vector $X$ of length 4 can be encoded into a 2-qubit state $\psi$ like 
\begin{align}
    X = 
\begin{pmatrix}
x_0 \\
x_1 \\ 
x_2 \\ 
x_3 
\end{pmatrix}
\mapsto \psi = \frac{1}{||X||} 
\begin{pmatrix}
x_0 \\
x_1 \\ 
x_2 \\ 
x_3 
\end{pmatrix}.
\end{align}

If a kernel method is used, the circuit only consists of an encoding layer (blue box in Fig.~\ref{fig_var_circuit}) and the measurement.
However, if a trainable kernel is applied, the circuit also contains a variational layer (green box in Fig.~\ref{fig_var_circuit}) and the general circuit layout is similar to the one for variational models.

A big challenge in QML is the selection of an appropriate design for the encoding and variational circuits, or for the kernel circuit, respectively. 
Different metrics were introduced for this reason, focusing, e.g. on the parameterized circuit's ability to approximate the distribution of unitaries according to the Haar measure for randomly sampled parameters \cite{Sim2019_expressibility}, the Fourier frequency spectrum \cite{Schuld2020_effect_of_encoding}, or the trainability of the circuit based on the variance of the cost function gradient \cite{Holmes2021_gradients}.

\subsection{Support Vector Regression}
Support Vector Regression (SVR) implements a generalized version of the more known Support Vector Machine (SVM), and was introduced by Drucker, Burges, Kaufman, Smola and Vapnik  in 1996 \cite{Drucker1996}.
The main difference to the SVM is that SVR finds an optimal hyperplane with an $\epsilon$-tube around it that maps the input features to a continuous output variable such that most points are \textit{inside} the tube.
Similar to the SVM, the SVR makes use of the kernel trick to map the data into a higher dimensional kernel space where linear regression of the data is easier.

The $\epsilon$-kernel-SVR solves the constrained quadratic optimization problem \eqref{eq_svr_quadr_probl}, which is explained in more detail in \cite{Vapnik1998}, and of which a nice graphic explanation is given in \cite{Khanna2015}.
$C$ is a regularization parameter, $\xi$ and $\xi^*$ are slack variables that determine how many outliers can be tolerated, and $w$ is a vector of weights determining a hyperplane in space.
$N$ is the number of training points, $x_i$ a single datum, $\varphi$ the transformation from feature to kernel space, and $2\epsilon$ is the width of the margin within which the loss of a data point is $0$.

The dual problem is formulated in \eqref{eq_svr_dual_problem} with Lagrange multipliers $ \alpha, \alpha^*$, and $N_{SV} \leq N$ is the number of support vectors. The kernel is defined as in \eqref{eq_kernel}.

\begin{equation}
\min_{w,\xi,\xi*} \frac{1}{2} w^T w + C \sum_{i=1}^{N} (\xi_i + \xi_i^* )
\label{eq_svr_quadr_probl}
\end{equation}
subject to
\begin{align*}
y_i - w^T \varphi(x_i) &\leq \epsilon + \xi_i^* \, &&i= 1 \dots N \\
w^T \varphi(x_i) - y_i &\leq \epsilon + \xi_i \, &&i=1 \dots N \\
\xi_i ,\:  \xi_i^* &\geq 0 \, &&i= 1 \dots N
\end{align*}

\begin{align}
 \max_{\alpha, \alpha^*} \: - \epsilon \sum_{i=1}^{N_{SV}} (\alpha_i + \alpha_i^*) + \sum_{i=1}^{N_{SV}} (\alpha_i^* - \alpha_i) y_i \label{eq_svr_dual_problem}  \\
- \frac{1}{2} \sum_{j=1}^{N_{SV}} \sum_{i=1}^{N_{SV}} (\alpha_i^* - \alpha_i)(\alpha_j^* - \alpha_j) k(x_i,x_j) \notag \\
\intertext{where}
 \alpha_i, \alpha_i^* \in [0,C], i=1,\dots,N_{SV}, \sum_{i=1}^{N_{SV}} (\alpha_i^*  \alpha_i) = 0 \notag 
\end{align}

\begin{equation}
    k(x_i, x_j) = \langle \varphi(x_i) , \varphi(x_j) \rangle \label{eq_kernel}
\end{equation}

\subsection{Support Vector Regression with Quantum Kernel}
\begin{figure}[htbp]
\centerline{\includegraphics[width=.5\textwidth]{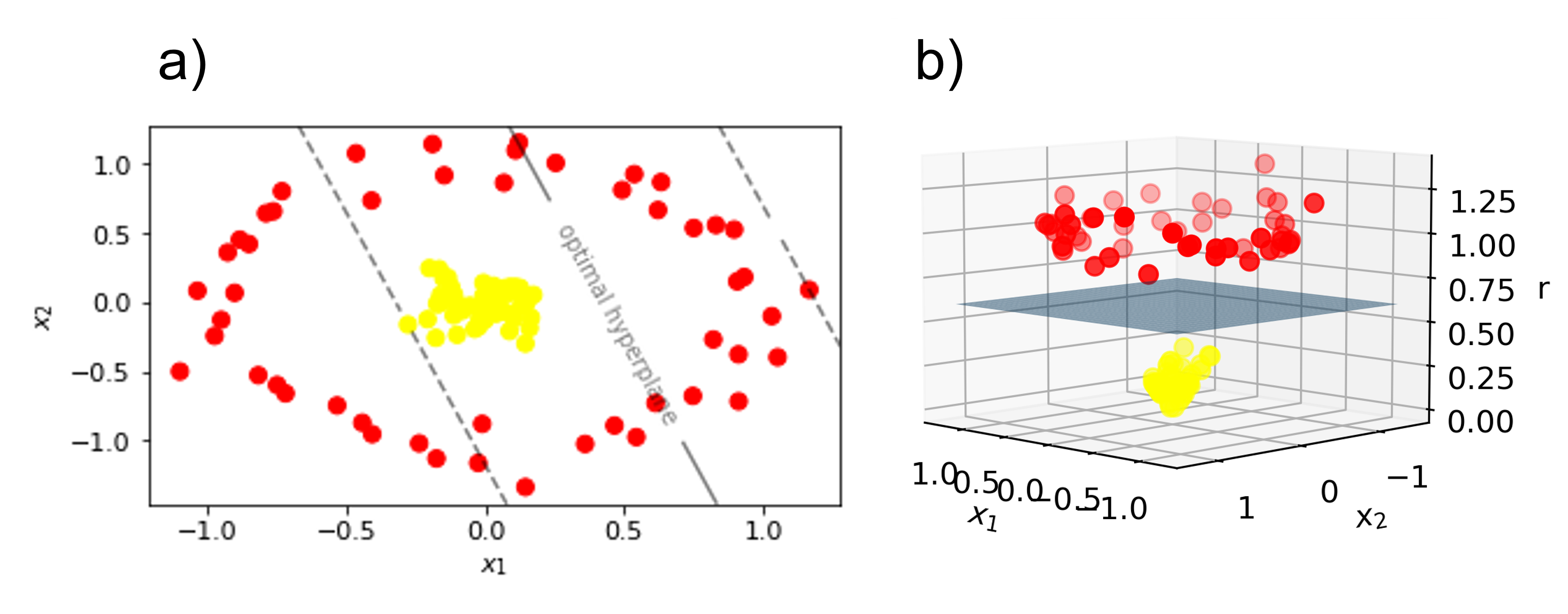}}
\caption{Example of the kernel trick for a Support Vector Machine.
a) original, two dimensional data with optimal separating hyperplane and margin.
The data cannot be separated by a hyperplane.
b) data after transformation into a third dimension which is the distance from center.
The data is now easily separable by a hyperplane.}
\label{fig_svr_kernel_trick}
\end{figure}

The advantage of using a kernel can be seen in Fig. \ref{fig_svr_kernel_trick}.
In a), a two dimensional data set is shown containing two classes, red and yellow, that cannot be separated by a hyperplane.
If we add the distance from origin as an additional dimension to the data, as it is done in Fig. \ref{fig_svr_kernel_trick} b), the two classes can be separated by a hyperplane.
In this case the transformation from feature to kernel space is achieved by calculating the distance from the origin for each datum.

Some kernels, like the example above, can easily be calculated on an ordinary computer.
Other kernels however, are hard to compute classically, whereas quantum computers have the potential to efficiently calculate them \cite{Schuld2021_qml_are_kernel}.

For the quantum kernel we utilize a quantum feature map to apply the kernel trick.
More precisely, the transformation in the quantum kernel is 
performed by a quantum feature map $\psi(x)$ that maps a datum $x$ to a Hilbert space.
The entries of the quantum kernel $K(x_i, x_j)$ are the fidelities or transition amplitudes between the states $\ket{\psi(x_i)}$ and $\ket{\psi(x_j)}$, which are the transformed feature vectors $x_i$ and $x_j$, respectively \cite{Havlicek2019, Glick2022}.

For the pure states $\ket{\psi(x_i)}$ and $\ket{\psi(x_j)}$, prepared by the Unitary $U$, the kernel is given by:
\begin{align}
     K(x_i,x_j) &= F(\psi_{x_i}, \psi_{x_j}) = \vert \braket{\psi_{x_i}|\psi_{x_j}} \vert ^2 \notag \\
     &= \vert \bra{0^{\otimes n}} U_{\psi_{x_i}}^\dag U_{\psi_{x_j}} \ket{0^{\otimes n}} \vert ^2 \label{eq_quantum_kernel}
\end{align}

The quantum kernel can be interpreted as a similarity measure between two transformed samples.
If $x_i = x_j$, i.e. $\psi_{x_i} = \psi_{x_j}$, the kernel equation \eqref{eq_quantum_kernel} gives:
\begin{align}
     K(x_i,x_j) &= K(x_i,x_i) = F(\psi_{x_i}, \psi_{x_i}) = \vert \braket{\psi_{x_i}|\psi_{x_i}} \vert ^2  \notag \\
     &= 1 \quad  \text{if} \; x_i = x_j
     \label{eq_quantum_kernel_same_state}
\end{align}

In the opposite case, if $\psi_{x_i}$ and $\psi_{x_j}$ are orthogonal, the kernel will evaluate to: 
\begin{align}
     K(x_i,x_j) &= F(\psi_{x_i}, \psi_{x_j}) = \vert \braket{\psi_{x_i}|\psi_{x_j}} \vert ^2 \notag \\
     &= 0 \quad  \text{if} \; \psi_{x_i} \perp \psi_{x_j}
     \label{eq_quantum_kernel_orth_state}
\end{align}

On a quantum computer, the kernel circuit in \eqref{eq_quantum_kernel} is prepared for each possible pair of train samples and the probability of measuring the all zero string in the $Z$-basis is an estimate of the fidelity of their corresponding encoded quantum states.

For predicting a new datum $p$, it is sufficient to estimate the kernel with all support vectors $s_i$ for $i = 1, \dots, N_{SV}$ which can make the evaluation of $K$ faster than during training.

As quantum kernel methods are linear models in the feature space, they can be efficiently evaluated if the inner products of feature vectors can be accessed \cite{Schuld2021_qml_are_kernel}.
Moreover, the optimal measurement for a kernel model on an $n$-qubit quantum computer is defined by $M \ll 2^{2n}$ degrees of freedom, whereas a variational model requires $\bigO(2^{2n})$ parameters to explore the whole feature space and be guaranteed to find the optimum model \cite{Schuld2021_qml_are_kernel}.
In addition, the optimisation problem \eqref{eq_svr_dual_problem} for a kernel model is convex for a convex loss function, which may help with trainability problems like barren plateaus \cite{Schuld2021_qml_are_kernel, McClean_2018_barren}.
On the other hand however,  quantum kernel estimation has unfavorable scaling for the number of samples.
During training it scales $\bigO(N_{\text{train}}^2)$ with the number of train samples, and during testing the scaling is $\bigO(N_{\text{SV}}N_{\text{test}})$.

\subsection{Autoencoder for Anomaly Detection}
\begin{figure}[htbp]
\centerline{\includegraphics[width=.5\textwidth]{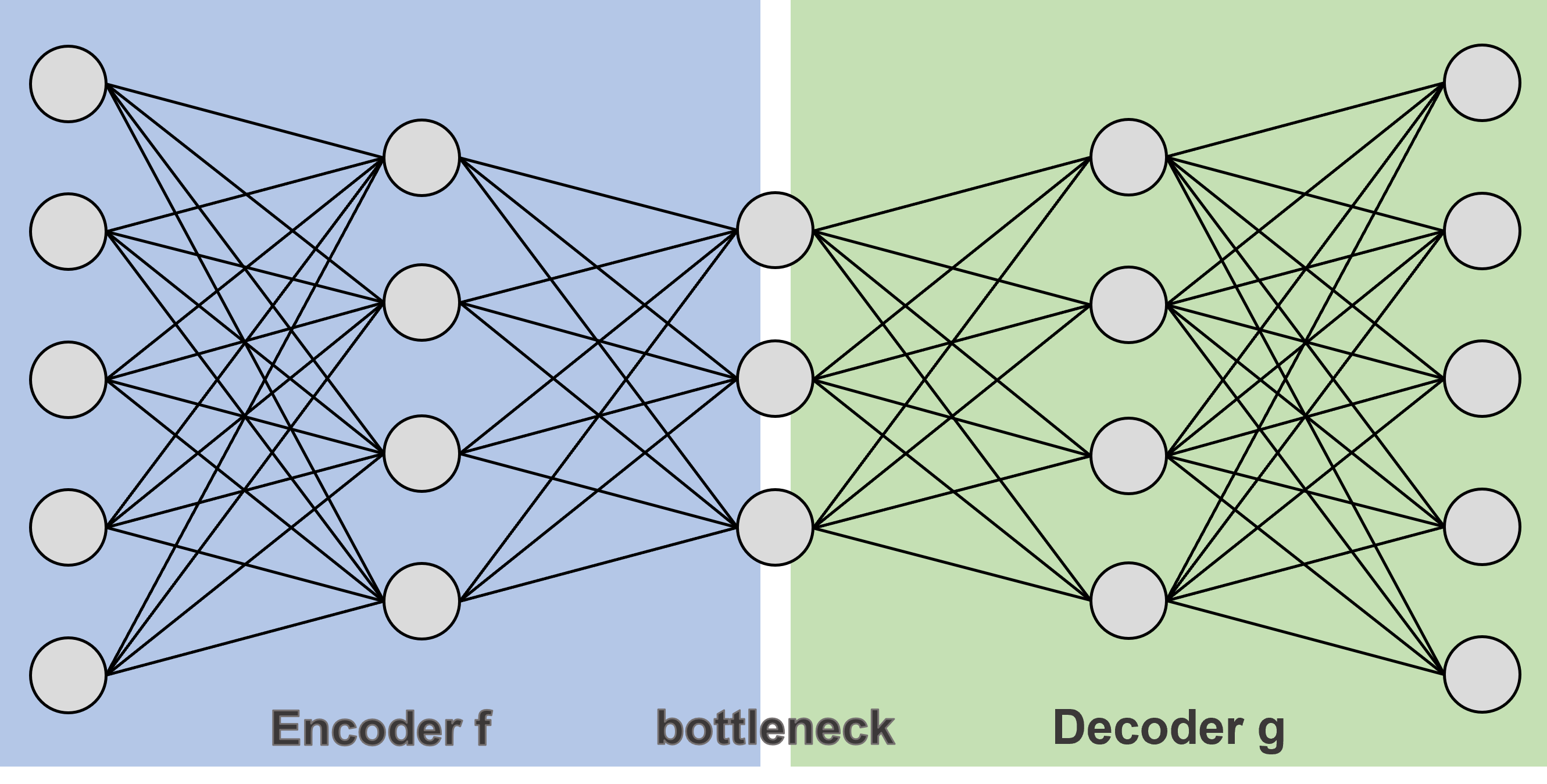}}
\caption{General architecture of an autoencoder consisting of encoder and decoder with the bottleneck layer.}
\label{fig_autoencoder_architecture}
\end{figure}

Autoencoders are a type of artificial neural network commonly utilized for unsupervised learning tasks, such as dimensionality reduction and data compression.
They are also popular reconstruction-based models for anomaly detection \cite{huang2022enhancing}.
The general architecture of an autoencoder can be seen in Fig.~\ref{fig_autoencoder_architecture}.
Autoencoders are composed of an encoder function, denoted as $f_\theta: \mathcal{X}\times\Theta \to \mathcal{F}$, which maps input data from the input space $\mathcal{X}$ to a lower-dimensional space $\mathcal{F}$, and a decoder function, denoted as $g_\theta: \mathcal{F}\times\Theta \to \mathcal{X}$, which maps the compressed representation back to the original input space $\mathcal{X}$.
In most cases $f_\theta$ and $g_\theta$ are realized via neural network architectures and parameterized by weights coming from some parameter space $\Theta$ and nonlinear activation function.
To achieve the embedding in the low-dimensional feature space, encoder and decoder architectures are usually symmetric around a common \emph{information bottleneck} layer with less units than input and output layer such that the model cannot just learn the identity function.

During training, the objective is to minimize the reconstruction loss, which is the difference between the input datum $x$ and the output $(g_\theta \circ f_\theta)(x;\theta)$ of the decoder applied to the encoded input relative to some norm $\Vert\cdot\Vert$:
\begin{equation}
    \min_{\theta} \Vert x-(g_\theta \circ f_\theta)(x)\Vert 
\end{equation}

The encoding process results in a bottleneck at the compressed representation $F$, where some information about the input data may be lost.
To achieve a small reconstruction loss, the encoder must learn to capture the relevant features of the data while discarding the irrelevant ones. This makes autoencoders well-suited for anomaly detection, as they can learn to effectively compress and restore normal data samples close to their original values. In contrast, the output of anomalous data samples would deviate significantly from their original input.

\subsection{Quantum Autoencoder for Anomaly Detection}
To use the quantum autoencoder for anomaly detection as introduced by Kottmann, Metz, Fraxanet and Baldelli \cite{Kottmann2021}, it is sufficient to implement the encoder only, which is a variational circuit consisting of a sequence of encoding and variational layers as shown in Fig. \ref{fig_var_circuit}. 
After encoding the data into the circuit, a subset of the qubits is measured.
The expectation values of these so-called \emph{"trash qubits"} are used to calculate a loss, which is minimized in an iterative process during training .
The idea behind this approach is that during training the parameters of the variational layers are adjusted such that the trash qubits become disentangled from the remaining qubits and stay at $\ket{0}$.
Hence, the trash qubits do not carry information about the training data, meaning the information is compressed onto the remaining, unmeasured qubits.
After the training is completed, the model can detect outliers, since they cannot be compressed as efficiently as the normal data. 
For anomalous data, the expectation value of the trash qubits will differ from the one for normal samples, resulting in an increased loss.

\section{Methods}
A SVR with quantum kernel (QSVR) for semisupervised anomaly detection was compared to a quantum autoencoder, a SVR with radial-basis-function kernel and a classical autoencoder.
The QSVR was benchmarked on a simple toy data set as well as on 10 popular real-world anomaly detection data sets.

\subsection{Quantum Support Vector Regression}
Schuld's \cite{Schuld2021_qml_are_kernel} finding that supervised QML models are kernel methods and that kernel-based training is guaranteed to find better or equally good quantum models than variational circuit training motivated the use of QSVR for anomaly detection.
The better scaling regarding the number of necessary parameters to find the ideal model and the convex loss for quantum kernel models supported our approach.

\begin{figure}[htbp]
\centerline{\includegraphics[width=.5\textwidth]{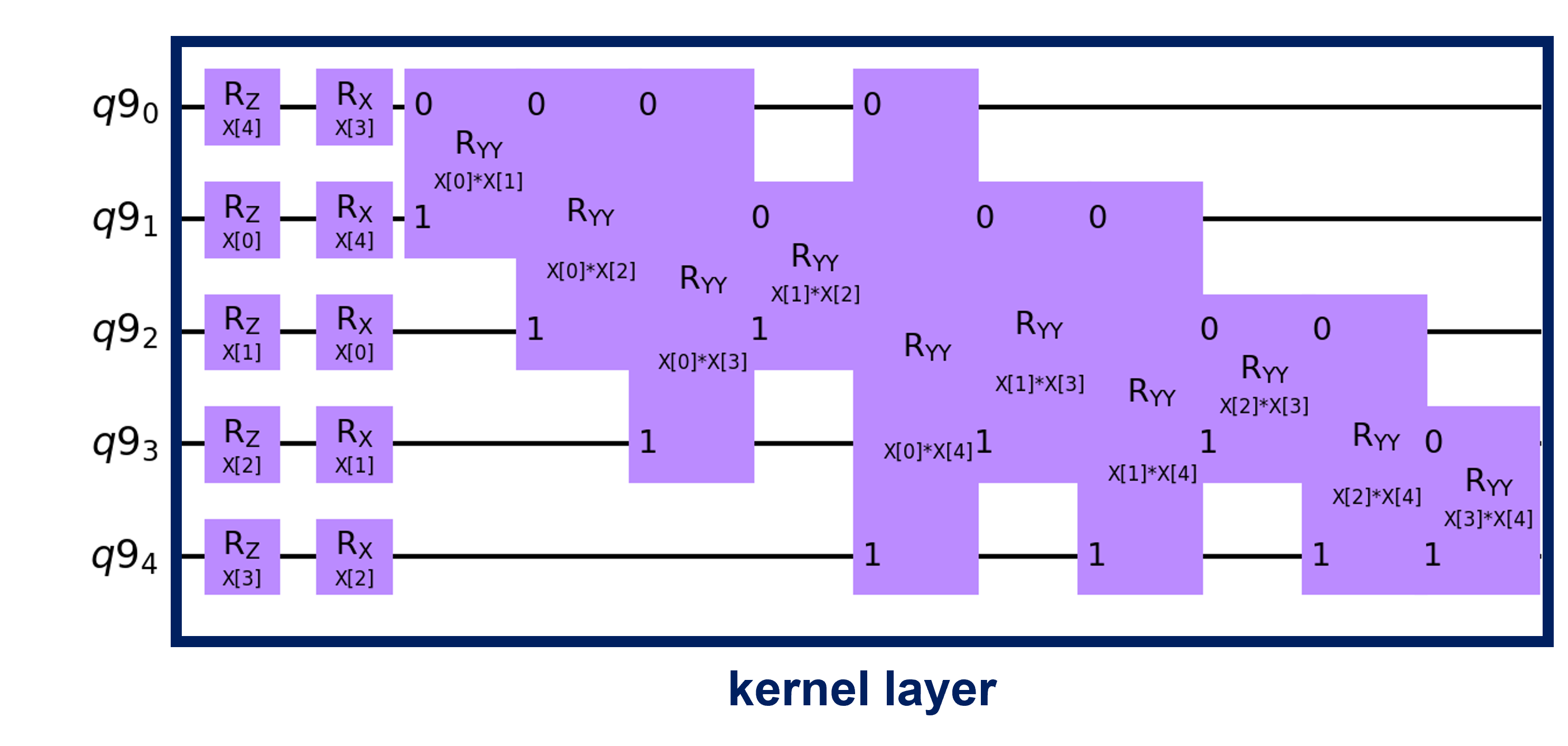}}
\caption{QSVR circuit. The kernel consists of a layer of $R_Z$ gates followed by a layer of $R_X$ gates and a layer of $R_{YY}$ gates that create full entanglement, i.e. each qubit is entangled to each other qubit through a $R_{YY}$ gate.
Each of the 5 features is encoded once into the $R_Z$ and once into the $R_X$ gates.
The parameters in the $R_{YY}$ gates are the product of two individual features.}
\label{fig_qsvr_circuit}
\end{figure}

Fig.~\ref{fig_qsvr_circuit} shows the circuit used to determine the kernel.
For the kernel, a slightly adapted version of the IQP embedding \cite{Havlicek2019} was chosen.
Two layers of single-qubit rotation gates were followed by a layer of two-qubit rotation gates.
The gates were selected to be $R_Z$, $R_X$ and $R_{YY}$.
Between the $R_Z$ and the $R_X$ layer, the order of the features was shifted so that each qubit had two different parameters for the $R_Z$ and $R_Y$ gates.
The parameters in the $R_{YY}$ gates were the products of two features.

In order to fit the output to the input dimension, an individual SVR with separate quantum kernel was trained for each of $N_{\text{feature}}$ feature dimensions, i.e. the model consisted of 5 QSVRs, each predicting one feature.
The reconstruction loss was the squared error between the input $y$ and the output $\hat{y}$ of the model.
This means our QSVR scales $\bigO(N_{\text{feature}}N_{\text{train}}^2)$ during training and $\mathcal{O}(N_{\text{feature}}N_{\text{SV}}N_{\text{test}})$ during testing.
The number of trainable parameters for a SVR is twice the number of train samples, however the trainable parameters for nonsupport vectors will be zero after training.
The number of support vectors depends on the data set.

\subsection{Quantum Autodencoder}
The quantum autoencoder is based on \cite{Kottmann2021}, where it has been employed to extract the phase diagram of a quantum system.
The model consists of the same three parts as a variational model, also seen in Fig. \ref{fig_var_circuit}.
\begin{enumerate}
    \item Encoding circuit
    \item Parameterized trainable circuit
    \item Measurement
\end{enumerate}
In this work, the model was adapted to nonquantum datasets by using an encoding layer that can handle classical data.
The same circuit design that was already used for the kernel was chosen for the encoding: 
two layers of single-qubit Pauli-rotation gates ($R_Z$, $R_X$) followed by a third layer of a two-qubit Pauli rotation gates ($R_{YY}$).
A similar design was selected for the trainable layer.
However, here the single qubit rotation gates were $R_Y$ and $R_X$, and the two-qubit rotation gate was $R_{ZZ}$.
The trainable layer was inserted once after the encoding layer and the parameters were initialized close to zero to avoid barren plateaus \cite{McClean_2018_barren, Cerezo_2021_variational_circuits, Holmes2021_gradients}.
Finally, the "trash qubits" were measured and the loss $\mathcal{L}$ was the sum of the probability $p$ of the trash qubits being one.
In other words, the loss was the average hamming weight of the trash bit strings.
In the case of two trash qubits, the loss is calculated as follows:
\begin{align}
    \mathcal{L} = p(\ket{01}) + p(\ket{10}) + 2 p(\ket{11})
\end{align}

Minimizing the loss during training equals compression of the input from 5 to 3 qubits, since the model was trained such that the trash qubits were always in the state $\ket{0}$ with high probability.
Once the model is trained, it is expected to efficiently compress normal data, but not anomalous data, resulting in a high loss for anomalous points.
The design of the trainable layer resulted in 20 trainable parameters.
The model was trained with batch size 1 and for 10 epochs. 

\subsection{Classical Support Vector Regression}
The classical SVR was used as the baseline for the QSVR.
It shares the same structure as the quantum model, except for the choice of kernel.
While the quantum model utilized a quantum kernel, the classical model used a radial basis function kernel.
As for the QSVR, the number of trainable parameters is twice the number of train samples and the number of nonzero parameters is twice the number of support vectors.

\subsection{Classical Autoencoder}
The encoder has three layers, the input layer with 5 neurons, one hidden layer with 4 neurons and the output layer with 3 neurons.
For the hidden and output layer the $\operatorname{ReLU}$ function was used as activation.
The decoder has the reversed architecture of the encoder, and used $\operatorname{ReLU}$ as activation function for the hidden layer and $\operatorname{tanh}$ for the output layer.
The total number of trainable parameters of this model is the sum of the trainable parameters of the encoder and decoder, i.e. $39 + 41 = 80$.
The model is trained for 500 epochs.

\subsection{Datasets and Preprocessing}
The model was benchmarked on 10 real-world data sets suitable for semisupervised learning and a simple toy data set.
An overview over the data sets is given in table \ref{tab_auc}.
The toy data set was constructed in the following way that ensures that it is separable.
First, the data was placed on a 5-dimensional hyperplane.
Second, the anomalous data was shifted perpendicular to the plane by a value randomly taken from one of the intervals $[-1.0, -0.4]$ or $[0.4, 1.0]$.

Dimensionality reduction of the real-world data sets was done using principal components analysis (PCA).
After performing  PCA, the 5 features with highest explained variance were selected and rescaled to the interval $[-1,1]$.
Finally, 30 samples from the normal class were chosen from the data set for training, and for testing 25 normal and 25 anomalous samples were selected.
After training, the anomaly classification threshold $\tau$ was set to be three times the mean training loss.

\subsection{Implementation details}
The kernel estimation for the QSVR was performed on the 27 qubit IBM Quantum System One "ibmq\_ehningen" with Falcon r5.11 architecture and no error mitigation techniques besides the defaults of the sampler primitive.
The simulations were performed on the qasm\_simulator.
The classical models were trained on a Nvidia DGX A100 system.
Qiskit \cite{Qiskit} was used as framework and preprocessing of the data was done with scikit-learn \cite{scikit-learn}, pandas \cite{pandas} and numpy \cite{numpy}. 

\section{Results and Discussion}
In this study, we evaluated the performance of our proposed Quantum Support Vector Regression (QSVR) model against the quantum baseline model Quantum Autoencoder (QAE, based on \cite{Kottmann2021}) and two classical baseline models, namely Classical Support Vector Regression (CSVR) and Classical Autoencoder (CAE).
The quantum models were simulated, however for two data sets our QSVR was additionally run on real hardware.
We benchmarked the models on 11 datasets: Credit Card Fraud (CC), Census, CoverT, DoH, EMNIST, FMNIST, KDD, MINST, Mammo, URL, and our constructed data set \emph{Toy}.
The models were evaluated based on AUC (Area under ROC curve), precision, recall, F1-score, and accuracy.

\subsection{Model performance}
\begin{figure*}[htbp]
\centerline{\includegraphics[width=\textwidth]{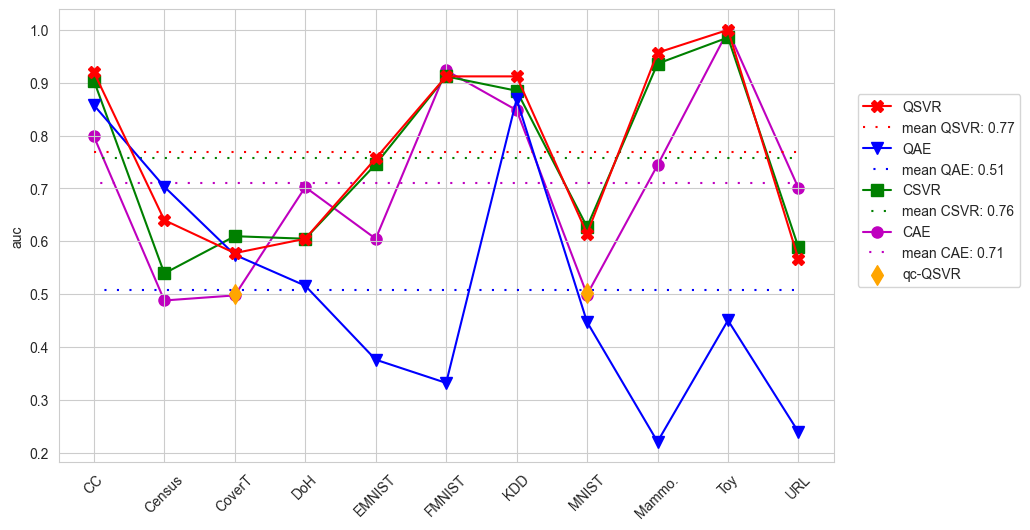}}
\caption{Plot of area under the curve (AUC) for the different models on the evaluated data sets.}
\label{fig_all_auc}
\end{figure*}

Special focus during evaluation is on the area under the ROC curve (AUC), a commonly used metric in anomaly detection that measures the trade-off between the true and false positive rate independent of a detection threshold.
An ideal model achieves an AUC of 1.0.
For a more comprehensive analysis, we report precision, recall, F1 score and accuracy as well.
Fig. \ref{fig_all_auc} shows the AUC for the individual data sets as well as the mean AUC over all data sets for each model.
Table \ref{tab_auc} reports the AUC, precision, recall, F1 score and accuracy for the models.
With a value of 0.77 our QSVR achieved the highest mean AUC of all models, closely followed by CSVR with 0.76.
The autoencoders performed worse on average, with CAE scoring a mean AUC of 0.71 and the QAE 0.51.
Considering the data sets individually, we see that the performance of the models varied strongly across the data sets. 
Our QSVR was the highest scoring model regarding the AUC on five data sets, and it was the best model in at least one of the five reported metrics for all but the URL data set.
The CSVR's individual performance was on a similar level as the QSVR's.
On CC, Census, MNIST, Mammo, and Toy, the QSVR achieved a (slightly) better AUC, whereas the CSVR scored (slightly) higher on CoverT and URL.
Compared to the QAE, which was selected as quantum baseline, our QSVR performed better on 9 out of 11 data sets, only being beaten on Census by an AUC of 0.06.
In comparison to the CAE, our QSVR performed better on 7 data sets, only being inferior to the CAE on DoH, FMNIST, and URL.

On the Toy data set, QSVR, CSVR and CAE performed well, reaching an AUC of 0.99 or above, which is not surprising since this data set was constructed such that it is easily separable.
The QAE, however, did not manage to pick up any information during training, resulting in an AUC of 0.45.
The Census, CoverT, DoH, MNIST and URL data sets were challenging for all models, as no model achieved an AUC above 0.70 on these data sets.

The CAE scored a recall of 0.0 and an undefined precision and F1 score on Census, CoverT, DoH, FMNIST, KDD, MNIST and URL.
The same applies to the QAE on CC, DoH, EMNIST, MNIST, Mammo and URL data sets.
This suggests that the models failed to learn from these data sets due to the anomaly score for  outliers falling below the threshold of three times the mean train loss and being too close to the anomaly score for the inliers.
Consequently the models could not identify any positives, neither true nor false, resulting in a recall of zero, as its numerator relies on the number of true positives.
Moreover, the denominator of precision, involving the sum of true and false positives, became zero, leading to an undefined value for that metric.
This also resulted in an undefined F1 score.

These results suggest that our QSVR, with simulated quantum kernel, can not only keep up with the other models, but on average even (slightly) outperformed all of them.
This result is promising, since our quantum kernel was not adjusted to the data set through quantum kernel alignment \cite{Glick2022}, a technique that fits the kernel to the ideal kernel for the data.

The observed performance of the models on the data sets is largely consistent with results from earlier studies conducted by other authors \cite{Schulze2022, Schulze2022_2}.
Differences in performance between the data sets are likely attributable to the varying complexities present in each data set.

Table \ref{tab_num_trainable_params} reports the number of nonzero model parameters for each model and data set.
This number was constant for QAE and CAE, whereas the SVRs had two nonzero model parameters, $\alpha$ and $\alpha^*$ (see eq. \ref{eq_svr_dual_problem}), for each support vector.
For the QSVR, on average 50 out of 300 parameters were nonzero, and for the CSVR 58 out of 300.
The QSVR had less or the same number of nonzero model parameters and support vectors compared to the CSVR for 8 of the 11 data sets.
Keeping in mind that the performance of QSVR and CSVR was similar but the QSVR needed on average less nonzero model parameters, we can conclude that in general using the quantum kernel lead to easier reconstruction of the data compared to the RBF kernel.

\begin{table}[htbp]
\centering
\caption{Number of nonzero model parameters for each model type and data set as well as its mean number of model parameters. The number in the parenthesis behind CSVR and QSVR gives the total number of model parameters.}
\label{tab_num_trainable_params}
\begin{tabular}{crrrr}
\toprule 
Data set & QSVR (300) & QAE & CSVR (300) & CAE \\
\midrule 
CC & 4 & 20 & 4 & 80 \\
Census & 50 & 20 & 48 & 80 \\
CoverT & 84 & 20 & 122 & 80 \\
DoH & 0 & 20 & 0 & 80 \\
EMNIST & 86 & 20 & 100 & 80 \\
FMNIST & 74 & 20 & 92 & 80 \\
KDD & 28 & 20 & 30 & 80 \\
MNIST & 100 & 20 & 114 & 80 \\
Mammo & 32 & 20 & 26 & 80 \\
Toy & 58 & 20 & 68 & 80 \\
URL & 38 & 20 & 36 & 80 \\
\textbf{Mean} & \textbf{50} & \textbf{20} & \textbf{58} & \textbf{80} \\
\bottomrule 
\end{tabular}
\end{table}

Finally, the kernel estimation of our QSVR was carried out on real hardware for the CoverT and MNIST data sets.
The AUC was 0.50 for both data sets, dramatically lower than that of our simulated QSVR, meaning our model had no advantage over random guessing.
However, it should be noted that no noise reduction or error mitigation techniques other than the defaults from the sampler primitive were used, and hence it should be possible to increase the performance of the model on real hardware. 

\begin{table*}[htbp]
\centering
\caption{Performance of models on selected data sets. The models with the highest metrics for each data set are highlighted.}\label{tab_auc}
\begin{tabular}{ccccccccccc}
\toprule 
Data set & Ref. & Normal & Anomalous & Model & AUC & Precision & Recall & F1 & Acc. \\
\midrule 
\textcolor{gray}{CC} & \textcolor{gray}{\cite{cc}} & \textcolor{gray}{Normal} & \textcolor{gray}{Anomalous} & QSVR (ours) & 0.92 & \textcolor{gray}{0.95} & \textcolor{gray}{0.72} & 0.82 & 0.84 \\
 &  &  &  & \textcolor{gray}{QAE} & \textcolor{gray}{0.86} & \textcolor{gray}{nan} & \textcolor{gray}{0.0} & \textcolor{gray}{nan} & \textcolor{gray}{0.5} \\
 &  &  &  & CSVR & \textcolor{gray}{0.90} & \textcolor{gray}{0.72} & 0.92 & \textcolor{gray}{0.81} & \textcolor{gray}{0.78} \\
 &  &  &  & CAE & \textcolor{gray}{0.80} & 1.0 & \textcolor{gray}{0.32} & \textcolor{gray}{0.48} & \textcolor{gray}{0.66} \\
\textcolor{gray}{Census} & \textcolor{gray}{\cite{census_income}} & \textcolor{gray}{$\leq$50k} & \textcolor{gray}{$>$50k} & QSVR (ours) & \textcolor{gray}{0.64} & 0.67 & \textcolor{gray}{0.16} & \textcolor{gray}{0.26} & 0.54 \\
 &  &  &  & QAE & 0.70 & \textcolor{gray}{0.5} & \textcolor{gray}{0.04} & \textcolor{gray}{0.07} & \textcolor{gray}{0.5} \\
 &  &  &  & CSVR & \textcolor{gray}{0.54} & \textcolor{gray}{0.52} & 0.44 & 0.48 & \textcolor{gray}{0.52} \\
 &  &  &  & \textcolor{gray}{CAE} & \textcolor{gray}{0.49} & \textcolor{gray}{nan} & \textcolor{gray}{0.0} & \textcolor{gray}{nan} & \textcolor{gray}{0.5} \\
\textcolor{gray}{CoverT} & \textcolor{gray}{\cite{forest_type}} & \textcolor{gray}{1-4} & \textcolor{gray}{5-7} & qc-QSVR (ours) & \textcolor{gray}{0.50} & \textcolor{gray}{0.5} & 1.0 & \textcolor{gray}{0.67} & \textcolor{gray}{0.5} \\
 &  &  &  & QSVR (ours) & \textcolor{gray}{0.58} & \textcolor{gray}{0.5} & 1.0 & \textcolor{gray}{0.67} & \textcolor{gray}{0.5} \\
 &  &  &  & QAE & \textcolor{gray}{0.57} & 1.0 & \textcolor{gray}{0.04} & \textcolor{gray}{0.08} & \textcolor{gray}{0.52} \\
 &  &  &  & CSVR & 0.61 & \textcolor{gray}{0.56} & \textcolor{gray}{0.92} & 0.7 & 0.6 \\
 &  &  &  & \textcolor{gray}{CAE} & \textcolor{gray}{0.50} & \textcolor{gray}{nan} & \textcolor{gray}{0.0} & \textcolor{gray}{nan} & \textcolor{gray}{0.5} \\
\textcolor{gray}{DoH} & \textcolor{gray}{\cite{DoH}} & \textcolor{gray}{Benign} & \textcolor{gray}{Malicious} & QSVR (ours) & \textcolor{gray}{0.60} & 0.67 & \textcolor{gray}{0.08} & \textcolor{gray}{0.14} & \textcolor{gray}{0.52} \\
 &  &  &  & \textcolor{gray}{QAE} & \textcolor{gray}{0.52} & \textcolor{gray}{nan} & \textcolor{gray}{0.0} & \textcolor{gray}{nan} & \textcolor{gray}{0.5} \\
 &  &  &  & CSVR & \textcolor{gray}{0.60} & \textcolor{gray}{0.56} & 0.6 & 0.58 & 0.56 \\
 &  &  &  & CAE & 0.70 & \textcolor{gray}{nan} & \textcolor{gray}{0.0} & \textcolor{gray}{nan} & \textcolor{gray}{0.5} \\
\textcolor{gray}{EMNIST} & \textcolor{gray}{\cite{emnist}} & \textcolor{gray}{A-M} & \textcolor{gray}{N-Z} & QSVR (ours) & 0.76 & \textcolor{gray}{0.88} & \textcolor{gray}{0.56} & \textcolor{gray}{0.68} & 0.74 \\
 &  &  &  & \textcolor{gray}{QAE} & \textcolor{gray}{0.38} & \textcolor{gray}{nan} & \textcolor{gray}{0.0} & \textcolor{gray}{nan} & \textcolor{gray}{0.5} \\
 &  &  &  & CSVR & \textcolor{gray}{0.75} & \textcolor{gray}{0.58} & 0.84 & 0.69 & \textcolor{gray}{0.62} \\
 &  &  &  & CAE & \textcolor{gray}{0.60} & 1.0 & \textcolor{gray}{0.04} & \textcolor{gray}{0.08} & \textcolor{gray}{0.52} \\
\textcolor{gray}{FMNIST} & \textcolor{gray}{\cite{fmnist}} & \textcolor{gray}{0-4} & \textcolor{gray}{5-9} & QSVR (ours) & \textcolor{gray}{0.91} & 0.96 & \textcolor{gray}{0.88} & 0.92 & 0.92 \\
 &  &  &  & \textcolor{gray}{QAE} & \textcolor{gray}{0.33} & \textcolor{gray}{0.33} & \textcolor{gray}{0.04} & \textcolor{gray}{0.07} & \textcolor{gray}{0.48} \\
 &  &  &  & CSVR & \textcolor{gray}{0.91} & \textcolor{gray}{0.64} & 0.92 & \textcolor{gray}{0.75} & \textcolor{gray}{0.7} \\
 &  &  &  & CAE & 0.92 & \textcolor{gray}{nan} & \textcolor{gray}{0.0} & \textcolor{gray}{nan} & \textcolor{gray}{0.5} \\
\textcolor{gray}{KDD} & \textcolor{gray}{\cite{KDD}} & \textcolor{gray}{Normal} & \textcolor{gray}{Anomalous} & QSVR (ours) & 0.91 & 1.0 & \textcolor{gray}{0.64} & 0.78 & 0.82 \\
 &  &  &  & \textcolor{gray}{QAE} & \textcolor{gray}{0.87} & \textcolor{gray}{0.92} & \textcolor{gray}{0.44} & \textcolor{gray}{0.59} & \textcolor{gray}{0.7} \\
 &  &  &  & CSVR & \textcolor{gray}{0.88} & \textcolor{gray}{0.68} & 0.84 & \textcolor{gray}{0.75} & \textcolor{gray}{0.72} \\
 &  &  &  & \textcolor{gray}{CAE} & \textcolor{gray}{0.85} & \textcolor{gray}{nan} & \textcolor{gray}{0.0} & \textcolor{gray}{nan} & \textcolor{gray}{0.5} \\
\textcolor{gray}{MNIST} & \textcolor{gray}{\cite{mnist}} & \textcolor{gray}{0-4} & \textcolor{gray}{5-9} & qc-QSVR (ours) & \textcolor{gray}{0.50} & \textcolor{gray}{0.51} & 1.0 & 0.68 & \textcolor{gray}{0.52} \\
 &  &  &  & QSVR (ours) & \textcolor{gray}{0.61} & \textcolor{gray}{0.51} & 1.0 & 0.68 & \textcolor{gray}{0.52} \\
 &  &  &  & \textcolor{gray}{QAE} & \textcolor{gray}{0.45} & \textcolor{gray}{nan} & \textcolor{gray}{0.0} & \textcolor{gray}{nan} & \textcolor{gray}{0.5} \\
 &  &  &  & CSVR & 0.63 & 0.54 & \textcolor{gray}{0.76} & \textcolor{gray}{0.63} & 0.56 \\
 &  &  &  & \textcolor{gray}{CAE} & \textcolor{gray}{0.50} & \textcolor{gray}{nan} & \textcolor{gray}{0.0} & \textcolor{gray}{nan} & \textcolor{gray}{0.5} \\
\textcolor{gray}{Mammo} & \textcolor{gray}{\cite{mammo}} & \textcolor{gray}{Normal} & \textcolor{gray}{Malignant} & QSVR (ours) & 0.96 & \textcolor{gray}{0.94} & \textcolor{gray}{0.64} & \textcolor{gray}{0.76} & \textcolor{gray}{0.8} \\
 &  &  &  & \textcolor{gray}{QAE} & \textcolor{gray}{0.22} & \textcolor{gray}{nan} & \textcolor{gray}{0.0} & \textcolor{gray}{nan} & \textcolor{gray}{0.5} \\
 &  &  &  & CSVR & \textcolor{gray}{0.94} & \textcolor{gray}{0.8} & 0.96 & 0.87 & 0.86 \\
 &  &  &  & CAE & \textcolor{gray}{0.74} & 1.0 & \textcolor{gray}{0.2} & \textcolor{gray}{0.33} & \textcolor{gray}{0.6} \\
\textcolor{gray}{Toy} & \textcolor{gray}{/} & \textcolor{gray}{Normal} & \textcolor{gray}{Anomalous} & QSVR (ours) & 1.00 & \textcolor{gray}{0.89} & 1.0 & 0.94 & 0.94 \\
 &  &  &  & \textcolor{gray}{QAE} & \textcolor{gray}{0.45} & \textcolor{gray}{0.6} & \textcolor{gray}{0.24} & \textcolor{gray}{0.34} & \textcolor{gray}{0.54} \\
 &  &  &  & CSVR & \textcolor{gray}{0.99} & \textcolor{gray}{0.74} & 1.0 & \textcolor{gray}{0.85} & \textcolor{gray}{0.82} \\
 &  &  &  & CAE & 1.00 & 1.0 & \textcolor{gray}{0.88} & 0.94 & 0.94 \\
\textcolor{gray}{URL} & \textcolor{gray}{\cite{URL}} & \textcolor{gray}{Benign} & \textcolor{gray}{Non-benign} & \textcolor{gray}{QSVR (ours)} & \textcolor{gray}{0.57} & \textcolor{gray}{0.33} & \textcolor{gray}{0.08} & \textcolor{gray}{0.13} & \textcolor{gray}{0.46} \\
 &  &  &  & \textcolor{gray}{QAE} & \textcolor{gray}{0.24} & \textcolor{gray}{nan} & \textcolor{gray}{0.0} & \textcolor{gray}{nan} & \textcolor{gray}{0.5} \\
 &  &  &  & CSVR & \textcolor{gray}{0.59} & 0.56 & 0.56 & 0.56 & 0.56 \\
 &  &  &  & CAE & 0.70 & \textcolor{gray}{nan} & \textcolor{gray}{0.0} & \textcolor{gray}{nan} & \textcolor{gray}{0.5} \\
\bottomrule 
\end{tabular}
\end{table*}

\section{Conclusion and Outlook}
A SVR with quantum kernel was evaluated in a semisupervised anomaly detection task on 11 data sets and compared to a quantum autoencoder, as well as to a SVR with RBF kernel and a classical autoencoder as classical baselines.
It was shown that our simulated QSVR outperformed the QAE on 9 of the 11 data sets and performed on the same level as the classical SVR with RBF kernel.
Future work should try to boost the performance of the QSVR using quantum kernel alignment \cite{Glick2022}.
Furthermore, we observed that the QSVR needed less support vectors and trainable parameters than the CSVR, showing that the data is easier to reconstruct in Hilbert space.
Finally, the kernel of our QSVR was estimated on real hardware for two data sets, unfortunately with a drastic decrease in performance.

It is a direction for future research to further investigate the robustness of the model against errors and evaluate the effect of different error mitigation techniques.
The goal should be to achieve satisfying performance of the model on a real device.
Moreover, a strong dependency of the models performances on the data sets were observed and the reason for that has to be evaluated as well.
Finally, scaling up the QSVR to accommodate higher dimensional input is essential, and the model needs to be reevaluated under these conditions.

\section*{Acknowledgment}
The research is part of the Munich Quantum Valley, which is supported by the Bavarian state government with funds from the Hightech Agenda Bayern Plus.

\bibliography{qad.bib}
\end{document}